\documentclass[preprint,12pt]{elsarticle}
\usepackage[T1]{fontenc}
\usepackage[utf8]{inputenc}
\usepackage{latexsym}
\usepackage[english]{babel}
\usepackage{indentfirst}
\usepackage{graphicx}
\usepackage{float}
\usepackage{lmodern}
\usepackage{bm}
\usepackage{geometry}
\usepackage{titlesec}
\usepackage{subfigure}
\usepackage{amsmath}
\usepackage{amssymb}
\usepackage{hyperref}
\usepackage{cancel}
\usepackage{natbib}
\usepackage{pifont}
\usepackage{tikz}
\usetikzlibrary{matrix,arrows,decorations,backgrounds,shapes,calc,fit}
\usetikzlibrary{fadings}
\usetikzlibrary{decorations.pathmorphing}
\usetikzlibrary{decorations.markings}
\usepackage{xcolor}
\usepackage{amsfonts}
\usepackage{slashed}
\tikzfading [name=radialfade, inner color=transparent!40, outer color=transparent!100]
\tikzfading [name=radialfade2, inner color=transparent!60, outer color=transparent!100]
\renewcommand{\d}{\ensuremath{\mathrm{d}}}

\newcommand{\x}{\bm{ x_{\perp}}}
\newcommand{\p}{\bm{ p_{\perp}}}
\newcommand{\ka}{\bm{ k_{\perp}}}

\newcommand{\intp}{\int \frac{\d^2 \p }{(2\pi)^2}}

\newcommand{\col}[1]{{\color{black} #1}}
\newcommand{\ma}[1]{{\mathcal{#1}}}

\definecolor{purp}{RGB}{1,1,1}
\tikzstyle{densely dashed}=          [dash pattern=on 1pt off 1pt]
\journal{Nuclear Physics A}
\begin{document}

\begin{frontmatter}



\title{Isotropization of the Quark Gluon Plasma}


\author{T. Epelbaum, F. Gelis}

\address{{\it  Institut de Physique Theorique (URA 2306 du CNRS)\\ CEA/DSM/Saclay 91191, Gif-sur-Yvette Cedex, France}}

\begin{abstract}
We report here recent analytical and numerical work on the theoretical treatment of the early stages of heavy ion collisions, that amounts to solving the classical Yang-Mills equations with fluctuating initial conditions.  Our numerical simulations suggest a fast isotropization of the pressure tensor of the system. This trend appears already for small values of the coupling constant $\alpha_s$. In addition, the system exhibits an anomalously small shear viscosity.

\end{abstract}


\end{frontmatter}

\section{Introduction} \label{s1}
During the past twelve years, heavy ion collisions at the RHIC and the LHC have established the formation of the Quark Gluon Plasma (QGP), a new state of matter in which the quarks and the gluons are deconfined. Moreover, the QGP seems to behave like a nearly perfect fluid, and to do so after a very short transient time: less than 1 fm/c \cite{Romat1}. This is assumed to be the case because relativistic hydrodynamical models with a very small value of the shear viscosity can successfully describe the experimental data \cite{LuzumR1}, and those models require a very early start of the hydrodynamical behavior in order to work. 
\vspace{0.1cm}

So far theoretical models have failed to predict such an early onset of hydrodynamical behavior. The framework that we have used in this work -- the Color Glass Condensate (CGC) effective theory \cite{GelisIJV1,MclerV1,MclerV2} -- even predicts at Leading Order (LO) a negative longitudinal pressure $P_L$ of the system at the initial time, with a value opposite to the energy density $\epsilon$ and the transverse pressure $P_T$ \cite{LappiM1}. In contrast, hydrodynamics requires a rather small anisotropy of the system. This apparent contradiction between theory and experiment has yet to receive a satisfactory answer, and has triggered numerous studies arguing that this fast hydrodynamization, along with the failure of QCD to predict it, may be a hint of a strongly coupled QGP \cite{HellerJW1}.
\vspace{0.1cm}

We will adopt another point of view in the present work. Since at high energies, $\alpha_s$ should be rather small for a non-zero time window, we will keep using a weakly coupled framework, and try to improve the CGC by taking into account higher order corrections. One can try first to take into account the next to leading order (NLO) contribution of the CGC. Unfortunately, the results are even worse: because of the presence of Weibel instabilities in the theory \cite{RomatV1},\cite{Mrowczynski}, the pressures increase exponentially and diverge as time goes to infinity. A major improvement was achieved in \cite{GelisLV3} where it was shown that one can resum all the fastest growing terms at each order of the perturbative expansion simply by evolving classically an initial condition formed by the superposition of the LO classical field and a Gaussian fluctuation whose spectrum can be obtained by a 1-loop calculation. The classical evolution with this fluctuating initial condition can be performed numerically with the help of a Monte-Carlo method, and is referred to as the classical statistical method. As a proof of concept, this approach was  tested for a scalar model in \cite{DusliEGV1}\cite{EpelbG1}\cite{DusliEGV2}. This model, although much simpler than QCD, shares some important features with the Yang-Mills theory: scale invariance at the classical level and most importantly the presence of instabilities (parametric resonance instead of Weibel instabilities). 

\vspace{0.1cm}

The only  theoretical ingredient missing up to now in the CGC framework was the NLO calculation that gives the correct spectrum of fluctuations. In this proceeding, we present this spectrum derived in \cite{EpelbG2} for the first time, and use it as an input of the classical statistical method in order to compute the time evolution of the energy-momentum tensor $T^{\mu\nu}$, to see whether or not the system isotropizes \cite{EpelbG3}.
\section{Initial conditions in the CGC effective theory}\label{s2}
\subsection{Background field: CGC at LO}
In the following, we will use the usual Fock-Scwhinger gauge choice\footnote{This gauge choice does not fix the gauge completely: residual gauge transformations that do not depend on $\tau$ are still allowed. But since we are only interested in gauge invariant observables, we do not need to fix it. Numerically we have checked that $T^{\mu\nu}$ is indeed independent of this residual gauge choice.} $A^{\tau}=0$. the letters $a,b,c$ denote color indices, while $i,j,$ denote transverse spatial indices. Let us recall the central result of \cite{KovnerMW1} that gives in the CGC framework and in the $(\tau,\eta,x,y)$ coordinate system the classical gauge fields at LO just above the forward light cone ($\tau_0=0^{+}$)
\begin{align}
\ma{A}^{i\col{a}}(\x)=\null&\alpha_1^{i\col{a}}(\x)+\alpha_2^{i\col{a}}(\x)\;,
&\ma{A}^{\eta\col{a}}(\x)=\null&\frac{ig}{2} \alpha_1^{i\col{ab}}(\x)\alpha_2^{i\col{b}}(\x)\;,\label{eq:mvfields}
\end{align}
where the fields $\alpha_{n=1,2}^{\col{a}}$ that depend on $(\tau,\eta,\x)$ are pure transverse gauge fields formed by the Wilson lines $\ma{U}_{n}$
\begin{align}
\alpha_{n}^{i\col{ab}}(\x)=\null&
\frac{i}{g}\ma{U}_{n}^{\col{ac}\dagger}(\x)\partial^i\ma{U}_{n}^{\col{cb}}(\x)\;,&
\ma{U}_{n}(\x)=\null&{\rm exp}\left(-ig\frac{1}{\nabla_\perp^{2}}\rho_n(\x)\right)\;.
\end{align}
The $\rho$ are random color sources for which we only have a probabilistic knowledge\footnote{The probability distribution function describing the $\rho$ depends on the energy scale through the JIMWLK equation \cite{IancuLM1}. The implementation of the energy dependence via the JIMWLK equation is beyond the scope of this work, and is therefore left for a later study. The McLerran-Venugopalan model was used instead}. They are of order $\frac{Q_s^{2}}{g}$, where $Q_s$ is the saturation scale.
\subsection{Spectrum of fluctuations: CGC at NLO}
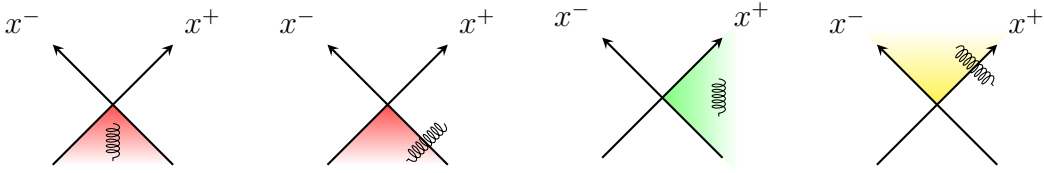
\begin{figure}[htbp]
\begin{align*}
&\begin{tikzpicture}[scale=0.5,gluonRouge/.style={decorate,decoration={coil,aspect=0.4pt,segment length=2pt},color=black},baseline=(current bounding box.center)]
\fill[color=red!70!white, path fading =south](-1.6,-1.6) -- (0,0) --(1.6,-1.6) ;
\draw[-stealth,thick] (1.6,-1.6) -- (-1.6,1.6) node [anchor=south east] {$x^-$};
\draw[-stealth,thick] (-1.6,-1.6) -- (1.6,1.6) node [anchor=south west] {$x^+$};
\draw[gluonRouge]  (0, -0.5) -- (0, -1.5);
\end{tikzpicture}&
&\begin{tikzpicture}[scale=0.5,gluonRouge/.style={decorate,decoration={coil,aspect=0.4pt,segment length=2pt},color=black},baseline=(current bounding box.center)]
\fill[color=red!70!white, path fading =south](-1.6,-1.6) -- (0,0) --(1.6,-1.6) ;
\draw[-stealth,thick] (1.6,-1.6) -- (-1.6,1.6) node [anchor=south east] {$x^-$};
\draw[-stealth,thick] (-1.6,-1.6) -- (1.6,1.6) node [anchor=south west] {$x^+$};
\draw[gluonRouge]  (1.5, -0.5) -- (0.5, -1.5);
\end{tikzpicture}
&\begin{tikzpicture}[scale=0.5,gluonRouge/.style={decorate,decoration={coil,aspect=0.4pt,segment length=2pt},color=black},baseline=(current bounding box.center)]
\fill[green!50!white, path fading = east](2,-2) --(2,-2) -- (0,0) -- (2,2) -- (2,-2) ;
\draw[-stealth,thick] (1.6,-1.6) -- (-1.6,1.6) node [anchor=south east] {$x^-$};
\draw[-stealth,thick] (-1.6,-1.6) -- (1.6,1.6) node [anchor=south west] {$x^+$};
\draw[gluonRouge]  (1.5, 0.5)--(1.5, -0.5) ;
\end{tikzpicture}&
&\begin{tikzpicture}[scale=0.5,gluonRouge/.style={decorate,decoration={coil,aspect=0.4pt,segment length=2pt},color=black},baseline=(current bounding box.center)]
\fill[yellow!80!white, path fading = north](-2,2) --(2,2) -- (0,0) ;
\draw[-stealth,thick] (1.6,-1.6) -- (-1.6,1.6) node [anchor=south east] {$x^-$};
\draw[-stealth,thick] (-1.6,-1.6) -- (1.6,1.6) node [anchor=south west] {$x^+$};
\draw[gluonRouge]  (0.5, 1.5)-- (1.5, 0.5);
\end{tikzpicture}
\end{align*}
\caption{\label{fig:smallf} The different steps that one has to perform in order to get the spectrum of fluctuations at $\tau_0>0$. From left to right: Evolution in the backward light-cone region on top of the vacuum, encountering the first nucleus, evolution on top of a (pure gauge) background field and encountering the second nucleus. A second contribution comes from encountering the two nuclei in opposite order.}
\end{figure}

One way to compute the NLO spectrum of the CGC is to consider plane waves in the remote past and make them evolve on top of the CGC LO classical field. The different steps of this evolution\footnote{The intermediate steps of this calculation were performed using the light-cone gauge condition advocated in \cite{BlaizY1}.} are illustrated in the figure \ref{fig:smallf}. Starting at $t=-\infty$ from a plane wave with momentum $(\ka,\nu)$, polarization $\lambda$ and color $c$:
\(
a^{\mu a}_{\ka \nu\lambda c}=\delta^{a}_c\epsilon^\mu_{{\bm k} \lambda}e^{ikx}\;,
\) 
(with the polarization vector $\epsilon^\mu_{{\bm k} \lambda}$ satisfying $k_\mu\epsilon^\mu_{{\bm k} \lambda}=0$ and $\epsilon^i_{{\bm k} \lambda}\epsilon^i_{{\bm k} \lambda'}=\delta_{\lambda\lambda'}$), the result obtained in \cite{EpelbG2} gives these small fluctuations at positive but small proper time $\tau_0\ll Q_s^{-1}$, after they have propagated on top of the two nuclei. The final result, reads\footnote{Where $\ma{D}^{iab}=\delta^{ab}\partial^{i}-ig\alpha^{iab}_1-ig\alpha^{iab}_2$.}
\begin{align}
a^{i a}_{\ka \nu\lambda c}
=\null&
F_{\ka\nu\lambda c}^{+,ia}+F_{\ka\nu\lambda c}^{-,ia}\;,&
a_{\ka \nu\lambda c}^{\eta a}
=\null&
\ma{D}^{iab}\; 
\Big(
\frac{F_{\ka\nu\lambda c}^{+,ib}}{2+i\nu}
-
\frac{F_{\ka\nu\lambda c}^{-,ib}}{2-i\nu}
\Big).
\label{eq:finalresult}
\end{align}
with 
\begin{align}
F_{\ka\nu\lambda c}^{\pm,ia}(\tau_0,\eta,\x)
=\null&
\Gamma(\mp i\nu)\,e^{\pm\frac{\nu\pi}{2}}
e^{i\nu\eta}\,\ma{U}^{ab\dagger}_{1,2 }(\x)\Big[\delta^{jk} -\frac{2k^j_\perp k^k_\perp}{k^2_\perp}\Big]\epsilon^k_{{\bm k} \lambda}\,\notag\\
&\times\intp\; e^{i\p\cdot\x}\;
\ma{\widetilde{U}}^{bc}_{1,2}(\p+\ka)
\quad\left(\frac{p_\perp^2\tau_0}{2k_\perp}\right)^{\pm i\nu}
\Big[\delta^{ij} -\frac{2p^i_\perp p^j_\perp}{p^2_\perp}\Big]
\; .
\end{align}
Those formulas provide the CGC NLO spectrum, at a proper time $Q_s\tau_0\ll 1$.
\subsection{The classical statistical method: resummed CGC}
In classical Yang-Mills (YM) simulations, it was argued in \cite{RomatV1} that Weibel instabilities can plague the numerical results if one adds rapidity dependent fluctuations on top of (\ref{eq:mvfields}). One way to circumvent this problem was proposed in \cite{GelisLV3}, where a resummed energy-momentum tensor was defined as follows 
\begin{align}
T^{\mu\nu}_{{\rm resum}} = \int \!\! \left[{\cal D} a\right]\,
F_0\left[a\right]\; T_{_{\rm LO}}^{\mu\nu} \left[\ma{A} + a\right] (x)\; , 
\label{eq:resum}
\end{align} 
where $F_0\left[a\right]$ is a Gaussian distribution of variance dictated by the spectrum of fluctuations (\ref{eq:finalresult}). This $T^{\mu\nu}_{\text{resum}}$ takes fully into account the first order quantum corrections (NLO), and a subset of every higher order corrections. Most importantly, it resums the terms that grow the fastest at each order of the perturbative expansion. Numerically, computing $T^{\mu\nu}_{{\rm resum}}$ amounts to solving the classical YM equations with a fluctuating initial condition of distribution $F_0$. This is the so-called classical-statistical method. When applied to scalar models \cite{DusliEGV1}\cite{EpelbG1}\cite{DusliEGV2}, this method was able to account for the macroscopic manifestations of a possible thermalization of the system: formation of an equation of state, and isotropization of the pressures. We have then applied (\ref{eq:resum}) to the YM case. 
\section{Implementation of the classical-statistical method}\label{s3}
\subsection{Numerical implementation}
Evaluating numerically (\ref{eq:resum}) can be done by a Monte-Carlo sampling of the initial condition formed by the sum of the classical background field (\ref{eq:mvfields}) and the NLO spectrum (\ref{eq:finalresult}) weighted by random gaussian coefficients
\begin{align}
A^{\mu\col{a}}(\tau_0,{\bm x}_\perp,\eta)=\null&\ma{A}^{\mu\col{a}}({\bm x}_\perp)+\sqrt{\frac{1}{2\pi V}}{\rm Re}
 \int_{\ka\nu}\sum_{\lambda   \col{c}}
 c_{\nu\ka\lambda\col{c}}\,a_{\nu\ka \lambda\col{c}}^{\mu\col{a}}(\tau_0,{\bm x}_\perp,\eta)\;,\label{eq:initspec}
\end{align}
$V$ being the lattice volume and $c_{\nu\ka\lambda}^{\col{c}}$ being random complex gaussian numbers of variance one 
\begin{align}
\left <c_{\nu\ka\lambda\col{c}}\,c_{\nu'\ka'\lambda'\col{d}*}\right >=\delta_{\nu\nu'}
\delta_{\ka\ka'}\delta_{\lambda\lambda'}\delta_{\col{cd}}\;.
\end{align}
One then just performs the time evolution of (\ref{eq:initspec}) through the classical Yang-Mills equations written in Hamiltonian formulation
\begin{align}
E^{\mu a}=\null& -\tau g^{\mu\nu} \partial_{\tau}A_{\nu a}\;,&
\partial_{\tau}E^{\mu a}=\null&\tau\,g^{\mu\nu} D^{\rho ab}F^{b}_{\rho\nu}\;,\label{eq:disceom}
\end{align}
with the \textsc{Gauss}'s law constraint
\(
D^{ab}_{\mu}E_{b}^\mu=\null0.
\)
In the process, one has to replace the gauge potential $A^{\mu}$ by link variables $U_\mu=e^{-ig a_\mu A_\mu}$ in order to exactly preserve gauge invariance on the lattice. The $a_\mu$ are the lattice spacings in the $\mu$ direction.
\subsection{Numerical results}
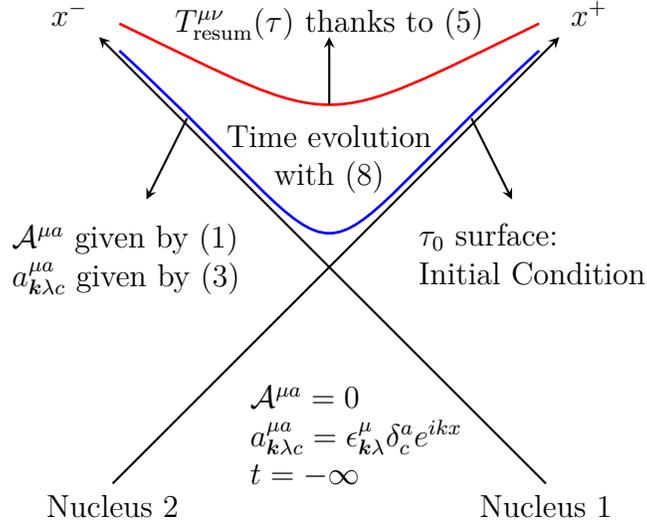
\begin{figure}[H]
\begin{center}
\begin{tikzpicture}[scale=1.8,gluonRouge/.style={decorate,decoration={coil,aspect=0.4pt,segment length=2pt},color=black},baseline=(current bounding box.center)]
\draw[blue,line width=1pt] (-1.55,1.6) .. controls (0.4,-0.2) and (-0.4,-0.2) .. (1.55,1.6);
\draw[-stealth,thick] (-1.05,1.1) -- (-1.35,0.5);
\draw[-stealth,thick] (1.05,1.1) -- (1.35,0.5);
\draw[-stealth,thick] (1.6,-1.6) node[anchor=north] {Nucleus 1} -- (-1.7,1.7) node [anchor=south east] {$x^-$};
\draw[-stealth,thick] (-1.6,-1.6) node[anchor=north] {Nucleus 2} -- (1.7,1.7) node [anchor=south west] {$x^+$};
\node[anchor=north,align=left] at (0.2, -0.8) {$
\mathcal{A}^{\mu a}=0$\\
$a_{{\bm k}\lambda c}^{\mu a}=\epsilon^{\mu}_{{\bm k}\lambda}\delta^{a}_c e^{ikx}
$\\
$t=-\infty$};
\node[anchor=north,align=left] at (-1.5, 0.4) {$
\mathcal{A}^{\mu a}$ given by (\ref{eq:mvfields})\\
$a_{{\bm k}\lambda c}^{\mu a}$  given by (\ref{eq:finalresult})};
\node[anchor=north,align=left] at (1.5, 0.4) {$\tau_0$ surface:\\ Initial Condition};
\node[align=center] at (0,0.8) {Time evolution\\ with (\ref{eq:disceom})};
\draw[red,line width=1pt] (-1.55,1.8) .. controls (0.1,1) and (-0.1,1) .. (1.55,1.8);
\node [] at (0,1.8) {$T^{\mu\nu}_{\text{resum}}(\tau)$ thanks to (\ref{eq:resum})};
\draw[-stealth,thick] (0,1.2) -- (0,1.7);
\end{tikzpicture}
\end{center}
\caption{\label{fig:shema} Schematic picture in light cone coordinate system of a heavy ion collision. We perform a classical statistical Yang-Mill simulation with the correct initial condition up to one loop on the blue surface $\tau_0=0^{+}$ in order to obtain $T^{\mu\nu}_{{\rm resum}}$ at later times.}
\end{figure}

\begin{figure}[H]
\begin{center}
\resizebox*{7.5cm}{!}{\includegraphics{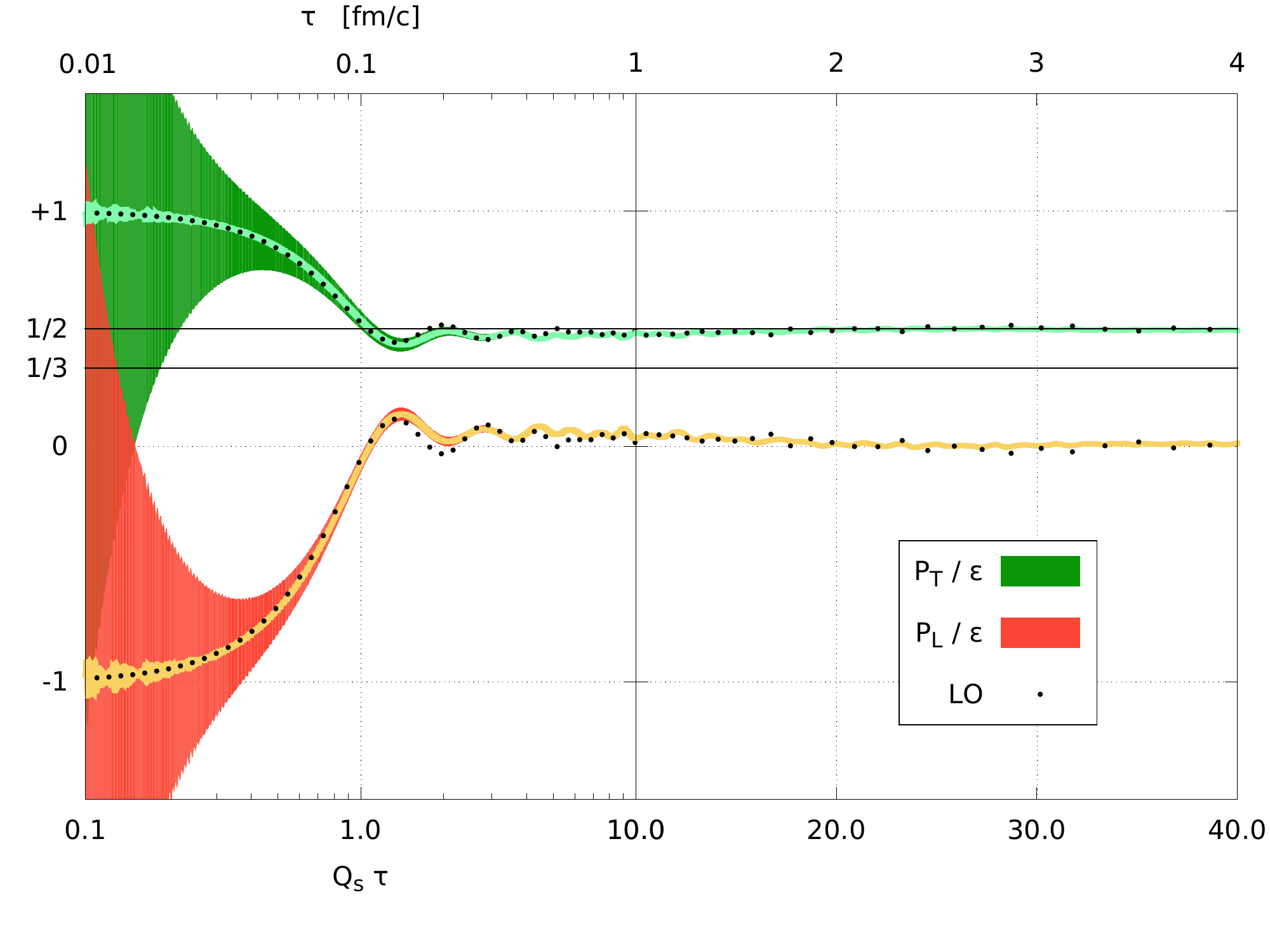}}\resizebox*{7.5cm}{!}{\includegraphics{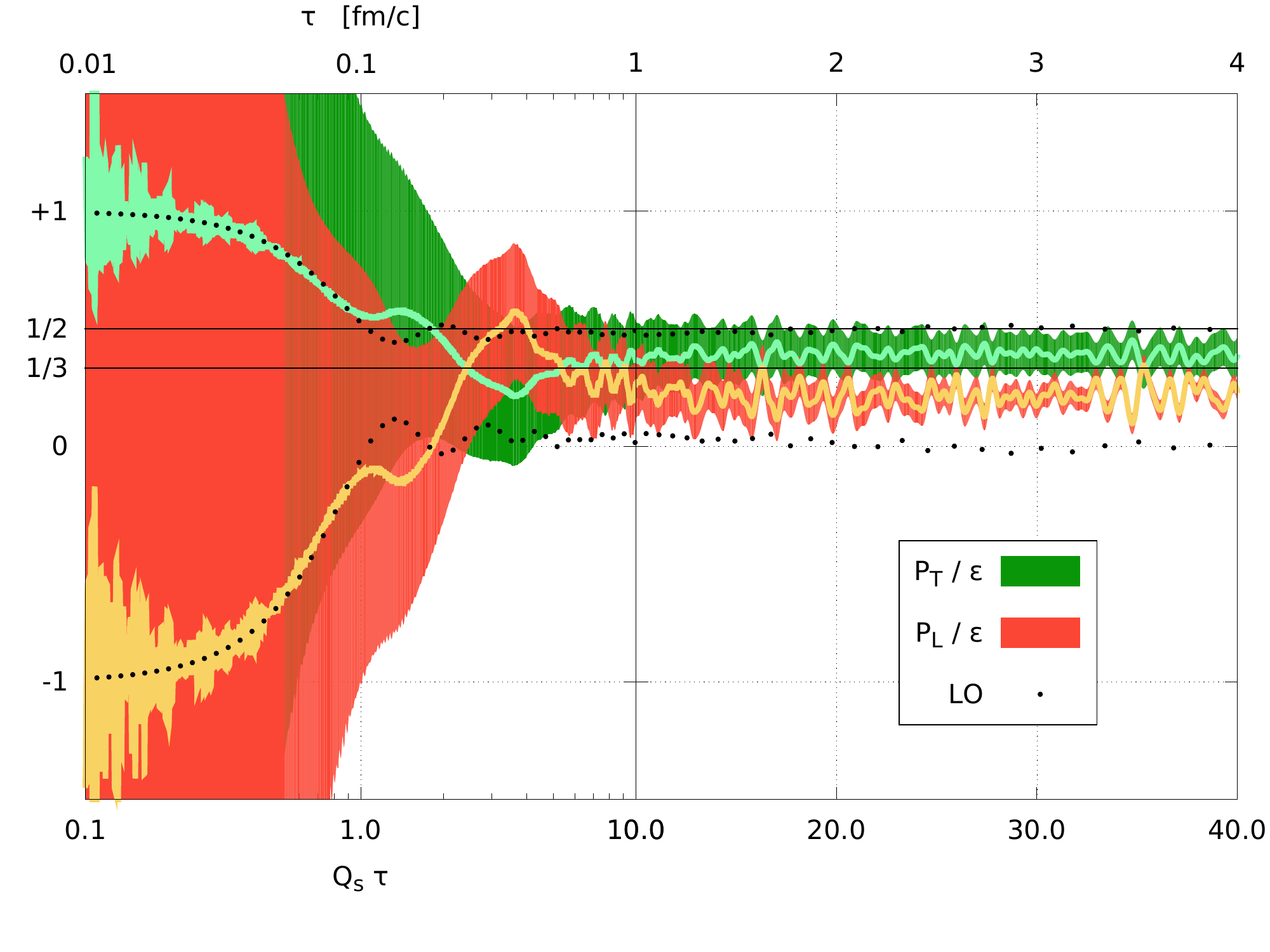}}
\end{center}
\vspace{-0.5cm} \caption{\label{fig:numres}$P_{_{T,L}}/\epsilon(\tau)$
  for $g=0.1$ ($\alpha_s=8\cdot 10^{-4}$, left plot) and $g=0.5$ ($\alpha_s=2\cdot 10^{-2}$, right plot). The bands indicate
  statistical errors. The dotted curves represent the LO result.} \vspace{-0.5cm}
\end{figure}

In the figure \ref{fig:shema}, we summarize the main steps of our computation\footnote{A related study was performed in \cite{BergeBSV1},\cite{BergeBSV2} at smaller coupling, starting at later times $Q_s \tau_0\approx 100$ with a different type of initial condition.}.
We compute at positive times the ratio $P_{_{T,L}}/\epsilon$, the components of $T^{\mu\nu}_{{\rm resum}}$ being averaged over the lattice volume and over the coefficients $c_{\nu\ka\lambda\col{c}}$ of the Monte-Carlo method. Our results for $g=0.1$ and $g=0.5$ are shown in the figure \ref{fig:numres}. The numerical computation has been performed on a $64\times 64\times 128$ lattice\footnote{Given the recent work performed in \cite{EpelGW}, let's notice here that a inherent cutoff dependance is present in the classical-statistical method. A systematical cutoff dependence study has yet to be done.}.
\vspace{0.2cm}
For $g=0.1$, the resummed result extracted from (\ref{eq:resum}) is very close to the pure LO result (black dots). This suggests that the Weibel instabilities do not influence much the dynamics at very early times for such a small value of the coupling. In contrast, $g=0.5$ sees an important qualitative change: the longitudinal pressure increases rapidly and the system reaches a fixed anisotropy (of the order of $40\%$) after a very short transient time of less than 1 fm/c. This is compatible with the very early start of the hydrodynamical behavior that was so far postulated, and observed here for the first time in a weakly coupled QCD framework. In addition, a very small value of the dimensionless ratio $\eta \epsilon^{-3/4}\sim 1$ -- roughly compatible with the values used in viscous hydrodynamical models -- can be obtained for $g=0.5$, by fitting the energy density with the help of a first order viscous hydrodynamical model $\epsilon=\epsilon_0 \tau^{-4/3}-2\eta \tau^{-1}$. This should be compared with the  LO perturbative value, approximatively equal to $300$. The conclusion is therefore that one does not need to invoke strong couplings in order to obtain a small value of the shear viscosity and to observe a fast isotropization of the QGP. This work is therefore an encouraging first step to reconcile the QCD treatment of the early stages of the QGP with the nearly perfect fluid picture that has emerged from experimental results.

\section{Acknowledgement}
This work is supported by the Agence Nationale de la Recherche project 11-BS04-015-01. Some of the
computations were performed with the resources provided by GENCI-CCRT
(project t2013056929).

\end{document}